\documentclass[conference]{IEEEtran} %!PN

\usepackage[dvips]{graphicx}
\usepackage{subcaption}
\usepackage{epsf}
\usepackage{epsfig}
\usepackage[noadjust]{cite}
\usepackage{amsmath}
\usepackage{url}
\usepackage[export]{adjustbox}
\usepackage{soul}
\usepackage[noend]{algpseudocode}
\usepackage{color}
\usepackage{tree-dvips}
\usepackage{float}
\usepackage{multirow}
\usepackage{comment}
\usepackage{soul}
\usepackage{booktabs}
\usepackage[dvipsnames]{xcolor}
\usepackage{siunitx}
\usepackage{footnote}
\usepackage{threeparttable}
\usepackage{pifont}
\usepackage[T1]{fontenc}
\usepackage[]{algorithm2e}

\def\BibTeX{{\rm B\kern-.05em{\sc i\kern-.025em b}\kern-.08em
    T\kern-.1667em\lower.7ex\hbox{E}\kern-.125emX}}

\newsavebox{\mybox}
\newlength{\mydepth}
\newlength{\myheight}

{\begin{lrbox}{\mybox}\begin{minipage}{\textwidth}}%
{\end{minipage}\end{lrbox}%
  \settodepth{\mydepth}{\usebox{\mybox}}%
  \settoheight{\myheight}{\usebox{\mybox}}%
  \addtolength{\myheight}{\mydepth}%
  \noindent\makebox[0pt]{\hspace{-20pt}\rule[-\mydepth]{1pt}{\myheight}}%
  \usebox{\mybox}}
  
\begin{document}

\title{On the (In)security of Approximate Computing Synthesis} 

\author{\IEEEauthorblockN{Sheikh Ariful Islam  } 
 \IEEEauthorblockA{Department of Computer Science and Engineering\\
 University of South Florida \\
 Tampa, FL 33620\\
 Email: \{sheikhariful\}@mail.usf.edu}}

\maketitle

\begin{abstract}

The broad landscape of new applications requires minimal hardware resources without any sacrifice in Quality-of-Results. Approximate Computing (AC) has emerged to meet the demands of data-rich applications. Although AC  applies techniques to improve the energy efficiency of error-tolerant applications at the cost of computational accuracy, new challenges in security threats of AC should be simultaneously addressed. In this paper, we  introduce the security vulnerability of the concurrent AC synthesis. We analyze the threat landscape and  provide a broader view of the attack and defense strategy. As a case study, we utilize AC synthesis technique to perform malicious modifications in the synthesized approximate netlist. Similarly, we provide a scalable defense framework for trustworthy AC synthesis. 

% The recent surge in hardware security is significant due to offshoring the proprietary Intellectual property (IP). One distinct dimension of the disruptive threat is  malicious logic insertion, also known as Hardware Trojan (HT). HT subverts the normal operations of a device stealthily. The diversity in HTs activation mechanisms and their location in  design brings no catch-all detection techniques.  In this paper, we propose to leverage principle features of social network analysis to security analysis of Register Transfer Level (RTL) designs against HT. The approach is based on investigating design properties, and it extends the current detection techniques. In particular, we perform both node- and graph-level analysis to determine the direct and indirect interactions between nets in a design. This technique helps not only in finding vulnerable  nets that can act as HT triggering signals but also their interactions to influence a particular net to act as HT payload signal. We experiment the technique on 420 combinational HT instances, and on average, we can detect both triggering and payload signals with accuracy up to  97.37\%. 

\end{abstract}

\section{Introduction}
\label{sec:Intro}

There is a growing number of ubiquitous embedded systems (e.g., Internet-of-Things) that have emerged as compelling platforms for complex applications. Application domains ranging from signal processing to machine learning are data-intensive. The potential of these applications is pervasive and widespread. The applications running on small geometries
cannot achieve the  quality of the results without incurring a high computational cost. Besides, it is difficult to tune design parameters once the design is in the field. Approximate Computing (AC) has evolved in the lower level of computing stack to address escalating challenges towards performance\footnote{Performance refers to power, area, timing, and energy of a system.}  efficiency. AC has created a compelling case towards efficiency beyond traditional low power techniques (e.g., DVFS, clock/voltage gating, etc.). AC leverages the strength of approximation present in arithmetic modules (e.g., adder \cite{Jiang:2015:CRE:2742060.2743760, 7167270, 8060326} and multiplier \cite{7372600, 5718826}). It offers a trade-off between efficiency and accuracy. In reality, it is a new computing paradigm that designers are currently considering throughout the pre-silicon  \cite{opencores}. 

Additionally, there have been significant efforts on software-level approximation techniques that run on top of deterministic hardware \cite{Agarwal2009UsingCP,7459495,6493641,7298218,8885006,8357321,8839352}. Even if AC is efficient by itself, security is misconstrued in the AC system that increases the consequences of security failures. In the light of recent security requirement for planet-scale IoT, we should integrate security mechanism into the systems built on AC to ensure information assurance.

Current security practice is oriented from the application-level to the bottom of the system (hardware). To envision trustworthy and energy-efficient complex systems, there are examples of attacks in the SW/HW boundary and mitigation strategy. Although the security patch can be easily integrated into unprotected software, vulnerabilities in the lower layer (hardware) are often difficult to predict due to its immutability and nonbypassability \cite{Lee:2012:HES:2382196.2382323}. Moreover, the increasing complexity of hardware tends to increase security vulnerability which lacks clear visibility. Further, the rising need for an energy-efficient system complicates the detection of security failures during the design process. In  light of the complexity and heterogeneity of the system, we should extend the security practice (attack and defense) for the end-to-end AC system. 

A large number of error metrics are present in the AC system to evaluate
the objective function (accuracy). For example, hamming distances, arithmetic difference, squared error are commonly used as error metrics. However, security requirements are absent both in Approximate Modules (AM) \cite{8105821} and Approximate Synthesis (AS) \cite{7533498, 6241596, Lee:2017:HSA:3130379.3130421}. Moreover, the security demand of the applications running on top of the end user's device (constructed with AM) may vary. Hence, the security model of the module and the synthesis technique can meet the requirements that apply. 

In this paper, we focus on the trustworthiness of approximate synthesis that should satisfy the real-time performance. In particular, we provide a comprehensive framework to embed malicious circuitry (also known as Hardware Trojan, HT) in an approximate module that has a shorter activation period and does not decrease the efficiency of the AC system. But this would violate  the integrity and trustworthiness of the approximate computing system. We perform such attacks on AS by using AM where some of AM are infected by HT while the rest of AM are HT-free. The module information (HT-free and/or HT-infected) is  unknown to the system integrator.
As the current AS tool does not discriminate between secure and non-secure AM, HT can be easily  retrofitted into the AC system. Moreover, given multi-objective requirements of the AC system, design space exploration is possible, which increases the complexity of HT detection. Hence, the current AS technique can create systems with satisfactory performance without any security guarantee. To enhance attack scalability, the HT in a set of deterministic modules\footnote{Deterministic modules are commonly used for realizing security critical function (e.g., S-box in AES).} can also be utilized with AM,  with no predictable performance degradation from deterministic subsystems.

Accordingly, a defender can employ easily verifiable security metrics to detect the presence of HT. As the error is inherent in the AC system, traditional HT detection approaches should be modified due to the large design space of the AC system. As the design constraints of AM is unknown to a defender, the detection technique also needs to be flexible (e.g., multi-voltage multi-mode analysis). Similarly, due to the large HT space, the defender might additionally be concerned with the security of deterministic modules or subsystems. Hence, for all extant AC systems that lack HT detection capabilities, we provide a robust path based detection technique satisfying the perceived accuracy requirements. To the best of authors' knowledge, this is the first work that quantifies security as a functional requirement which is consequential as regular performance. The novelty and contributions of the proposed approach are:
\begin{itemize}
    \item scalable and untrustworthy approximate synthesis framework 
    to include out-of-spec components in approximate netlist while maintaining pre-defined accuracy.
    \item comprehensive detection framework using error and path profiling considering the presence of a wide range of  HT instances in the approximate netlist.
    \item easy integration of both attack and defense framework into current synthesis flow. 
\end{itemize}

% While it is true that the published works are heavily invested towards narrow focus (energy efficiency), we quantify security as a functional requirement which is consequential as regular performance. Security failure in any way can compromise even narrow objectives. 

The rest of the paper is organized as follows. Section \ref{sec:back} provides an overview of related works on the security of approximate computing. Sections \ref{sec:threat} and \ref{sec:problem} describe the attack model and related definitions. Section \ref{sec:proposed} proposes the attack and detection mechanism of malicious insertions respectively.
Section \ref{sec:result} presents the detection rate without any knowledge of the golden design. Section \ref{sec:conclude} draws the conclusion followed by future work.

\section{Background and Related Work}
\label{sec:back}

In this section, we present the current state of security practice in Approximate Computing, although significant improvement in modeling, error analysis, functional approximation, and CAD have been recognized. As our focus in this paper is to analyze the security vulnerability of AC synthesis, we present the ongoing works that affect the overall trustworthiness of AC.

Regazzoni {\em et al.} \cite{Regazzoni:2018:SDS:3240765.3243497} provide an in-depth study about the possible hardware security for the AC system. Although the authors mention the potential of security failures, the techniques to address these violations are not addressed. Moreover, there is no mention of security enhancement at a higher abstraction level. Yellu {\em et al.} \cite{Yellu:2019:STA:3299874.3319453} mention some progress in security threats of four broad domains, namely, circuit, storage, software, and system for AC. No attempt has been made to define security objectives during synthesis to ensure scalable trustworthiness during pre- and post-silicon AC system. Moreover, many security gaps still exist due to fundamental reusable requirements during any synthesis technique.

Venkataramani {\em et al.} \cite{6241596} propose technique as to how to embed regular Observability Don't Care (ODC) during AC synthesis, which can simplify the Boolean equations. The method reduces the original variables set that do not directly influence the primary outputs. However, techniques remain for inserting Hardware Trojan in Register Transfer Level (RTL) don't care condition \cite{7342387}. Nepal {\em et al.} \cite{7533498} presented Automated Behavioral Approximate Circuit Synthesis (ABACUS) of RTL description that employs widely used NSGA-II \cite{996017} to obtain Pareto frontiers of AC system. Thorough optimizations of Abstract Syntax Tree (AST) from RTL design are utilized to enhance the accuracy of AC circuits with no incorporation of trustworthiness.  Lee {\em et al.} \cite{Lee:2017:HSA:3130379.3130421} presented a high-level synthesis (HLS) framework for approximating loop-based program behavior. Although HLS provides a higher abstraction for architectural synthesis,  the authors did not mention any synthesis modification towards approximate trustworthiness.

Unlike previous techniques, our technique (a) does consider any attribute of AM
that can be exploited to introduce any modifications into existing AC synthesis and (b) presents an effective and scalable approach to detect any such modifications early on during the hardware design life cycle (HDLC).

\section{Threat Model}
\label{sec:threat}

We assume an attacker can control the entire life cycle of the AC system if h/she can include the stealthy behavior into legacy reusable hardware components (e.g., arithmetic modules).  Hence, an attacker would modify the specified function with less clear untrusted properties for various attack objectives.  Moreover, the fundamental improvements on Commercial Off The Shelf (COTS) intellectual property with no security  verification makes security assurance of AC synthesis questionable. During pre-silicon, third-party approximate IP vendors can modify parts of the design and sell to the particular IP buyer where the buyer may be fabless design house. Even if these modifications are  visible during functional verification, they are stealthy during trust verification of the composite system. Although the AC system will perform satisfactorily despite the mere presence of untrusted components, the precise specifications of undesired properties (e.g., small change in design) may leak secret information (e.g., key) and synthesis configurations to help in overbuilding, introduce incorrect functionality during rare triggering event, cause early failure of the device, etc.

% The potential threats of HT are often aligned with the attackers' objectives that include but not limited to the leaking secret information (e.g., key) and synthesis configurations to help in overbuilding, introducing incorrect functionality during rare triggering event, early failure of the device, etc.

% We assume the potential inclusion of out-of-spec components into the legacy design can happen during pre- and post-silicon. In pre-silicon, third-party IP vendors can modify parts of the design and sell to the particular IP buyer where the buyer may be fabless design house. Similarly, a system integrator can act as an insider attacker where he has access to design internals and can subvert any IP cores included in System-on-Chip (SoC). Even the IP core is encrypted, as a valid and regular user, the integrator would have the correct key to unlock the core. During post-silicon, an attacker in the foundry would have access to GDSII file which he can reverse engineer to insert HT by changing process parameters (e.g., lithographic mask).

\section{Problem Statement}
\label{sec:problem}

Approximate modules constitute a significant source of security vulnerability in approximate computing synthesis. Two fundamental properties, namely, error and power, are inherent in AM and  can be exploited during hardware synthesis to perform malicious  changes. The changes should appear with the same likelihood as a regular bug in hardware, be extensible to operate during the unsafe condition and have a certain degree of connectivity to appear as payload at design outputs. On the contrary, knowing and understanding the design for security early in HDLC can reduce the related risks and testing of the synthesized netlist. 

If we denote approximate module as AM$^{j}_{i}$ with error (E$^{j}_{i}$) and power (P$^{j}_{i}$), an attacker objective is to modify AM$^{j'}_{i'}$ with new error ($\delta$E$^{j'}_{i'}$) and power ($\delta$P$^{j'}_{i'}$). Here $i$ denotes different architectures (e.g., 1- and 2-bit approximate adder) of $j$ type operation (e.g., adder, multiplier). The required changes in error and power are dictated by the synthesis objectives that AC can tolerate on the  approximate gate-level netlist. If the synthesized netlist can encompass error (E') and power (P'), the following two constraints must satisfy:

\begin{equation}
   E' - \sum_{j=1} \sum_{i=1} E^{j}_{i} < \delta E 
\end{equation}
   
\begin{equation}
   P' - \sum_{j=1} \sum_{i=1} P^{j}_{i} < \delta P
\end{equation}

Note that, all AMs can not satisfy the above constraints. Hence, an attacker would resort to iterative approach and develop a non-unified technique. On the contrary, a defender should carefully infer the distribution of error ($\delta$E) and power ($\delta$P) to detect the presence of modified AMs.

\section{Proposed Attack and Detection Technique}
\label{sec:proposed}

\subsection{Malicious insertion in AC synthesis}
\label{sec:attack}

\begin{figure}[h]
\centering
\includegraphics[width=\columnwidth]{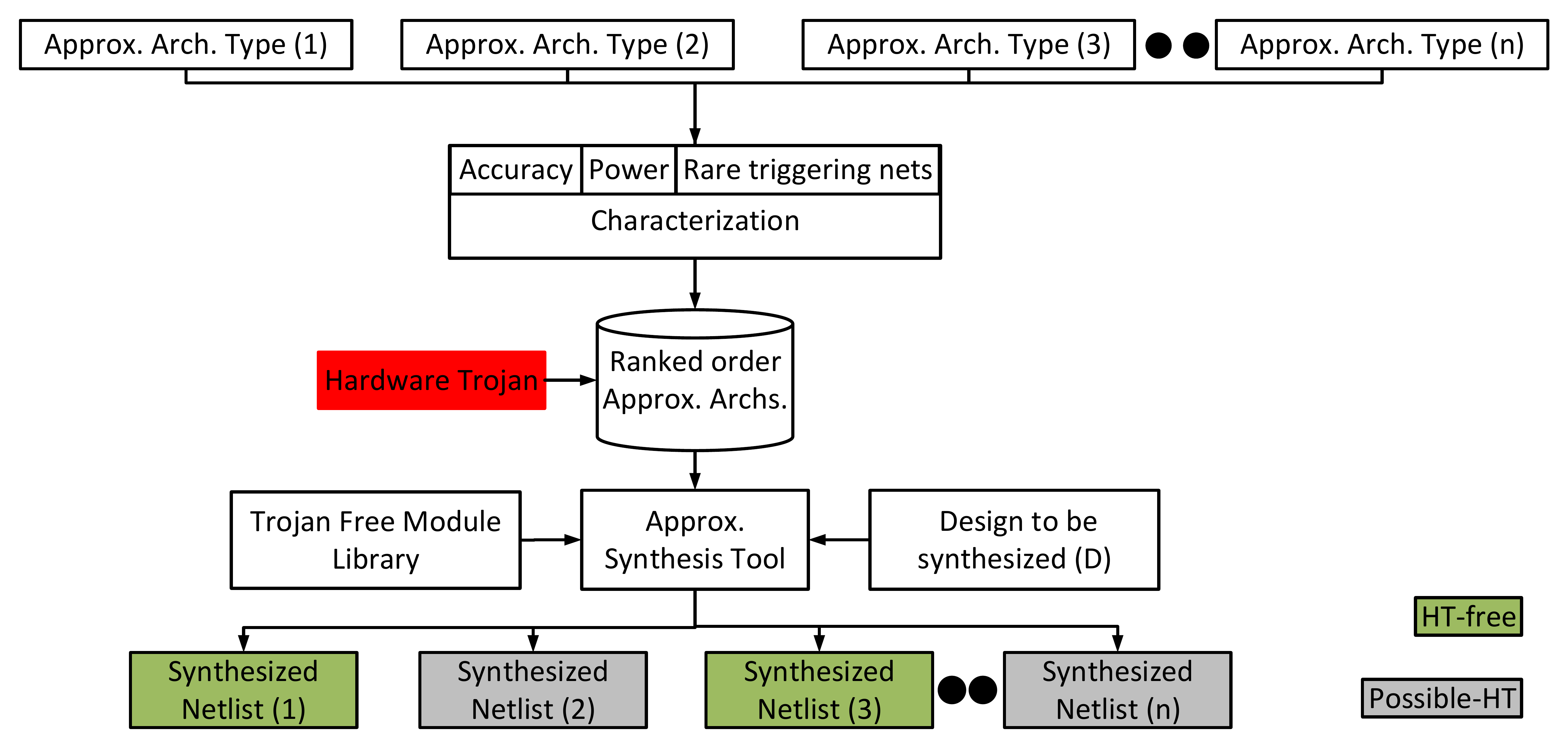}
\caption{HT insertion framework in AC synthesis flow (bottom-up approach).}
\label{fig:insertion}
\end{figure}

Current threats on hardware are often focused on the entity that does not participate in HDLC. For example, an attacker in manufacturing and test can embed untrustworthy components in the design. However, opportunities exist for malevolent actors during pre-silicon to exercise high-impact damage and leaks. 
An attacker can systematically evaluate the AM to gauge its security threats in the future which makes minimal use of AM. We show such an attack framework in Fig. \ref{fig:insertion}. The description of the framework is given as follows. 

Given access to approximate module library (architectures) and synthesis tool, an attacker would exploit the library and tool while still meeting the specified critical requirements of approximate design objectives. For example, approximate adder  and multiplier \cite{7167270} architectures are publicly available. First, we characterize each module architectures in terms of the objective function (accuracy, power, and rare triggering nets). Then, we perform functional simulation to  understand whether a particular
architecture can be maliciously attacked. To enforce approximate parameters (accuracy and power) while making the module vulnerable, we change the traditional design objectives of approximate synthesis as follows:
\begin{equation}
 Cost = W_{a,p} * (Accuracy + Power)\\ 
		+ W_{r} * Rare\_nets
\label{eq:leak_delay_cost}
\end{equation}

where, W$_{a,p}$  denotes the combined weights of accuracy and power, and  W$_{r}$ indicates the rare switching nets available within a module architecture.  We keep priority weights (W$_{a,p}$ and W$_{r}$) to 0.5. We perform an independent assessment of each module to determine 
the suitability of malicious modifications. Then, we rank the architectures of a particular module type (adder or multiplier) that show realistic adversarial behavior while meeting approximation criteria. We impose the following constraints while choosing a module for HT insertion:
\begin{itemize}
    \item the module showing higher error (less accuracy) is more susceptible to malicious modification.
    \item perform retiming and/or relaxing the paths that show timing error due to approximation and HT.
\end{itemize}

% The ranked order architectures of a particular module type (adder or multiplier) indicate the regular system objectives with realistic adversarial behavior. An attacker would consider the following constraints 

As there lacks standard threat infrastructure, we follow the automated HT insertion framework \cite{8342270} to measure the success of various HTs. Another challenge lies in checking the HT infected design
for attack success, which can be solved by using SCOAP \cite{1585245} for measuring controllability and observability. To broaden the attack surface and search time, we also provide HT-free modules to 
the approximate synthesis tool. Depending on synthesis tool configurations, we generate different synthesized netlists of the same functionality and pass it to the lower level (e.g., layout-level).  The generated netlist does not include any information related to the vulnerability of design.

\subsection{Detection of malicious modification in AC synthesis}
\label{sec:defense}

While approximate systems have seen significant innovation recently, critical issues, namely system-wide security vulnerability detection, remain unanswered.
Due to the large design space of the synthesized netlist of approximate systems, there are substantial new challenges to detect any malicious modification. Firstly, the  functionality of critical function (e.g., encryption) should not be fault-tolerant; hence, a designer typically avoids approximating such function. However, fault-tolerant applications (e.g., image processing) send/receive signals to/from critical function (e.g., biometric authentication). If the modifications are performed in fault-tolerant design (e.g., filter), the potential consequences are disastrous. Secondly, the vast optimization opportunities during approximate synthesis leave the current detection techniques only to the existing attack surface. Thirdly, approximate computing leverages cost-efficient data movement across different datapath units. Hence, datapath components exhibit a higher potential for intentional modifications. Finally, the security protocol must be able to handle the heterogeneity of accuracy and power objectives of underlying approximate modules at multiple scales. 

The security of the approximate system depends on the security of approximate modules plus the deterministic modules.  To understand the security vulnerability, it is essential to consider the conditions when any approximate module would be a rogue agent. 

\begin{figure}[h]
\centering
\includegraphics[width=\columnwidth]{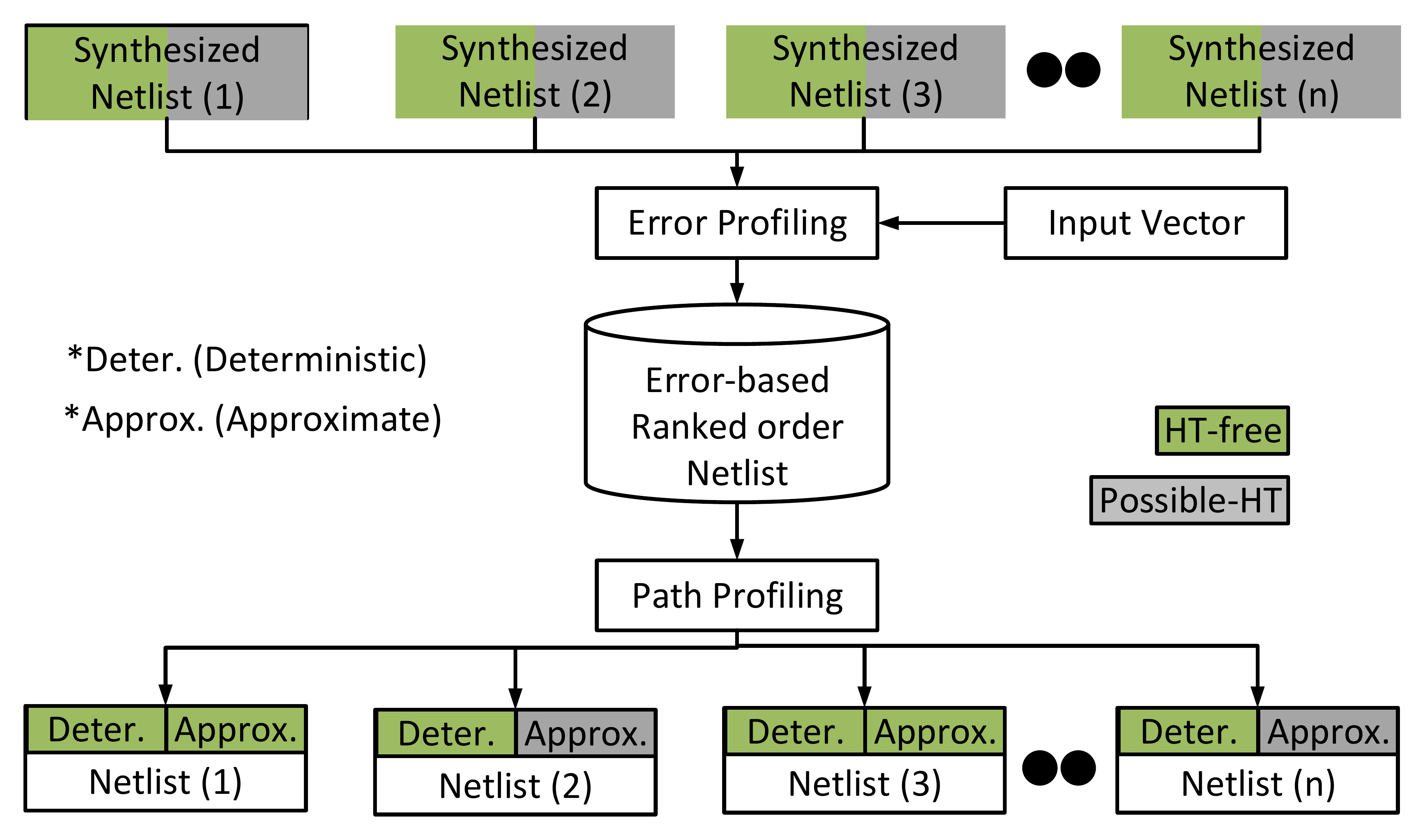}
\caption{HT detection framework in AC synthesis flow (top-down approach).}
\label{fig:detection}
\end{figure}

Consider the synthesized gate-level netlists from approximate synthesis tool (e.g., \cite{7533498}) or industry-standard tool (e.g., Design Compiler) in Fig. \ref{fig:detection}. To a defender, he does not have access to a golden netlist or approximate modules. Hence, he has to find a ``provably secure and energy-efficient'' design using a systematic approach. The search is more expensive if approximate computing circuits are domain-specific and deployed in IoT infrastructure. Hence, ad-hoc security procedures are too expensive to localize  malicious modification. While the analysis of design behavior is relatively easy for energy efficiency, the same is not true for security analysis. A defender then requires the formal treatment of energy efficiency of approximate computing to reason about the security vulnerability.  

Given $n$ netlists where the HT is carefully crafted for some netlists, a defender has to find the netlist where HT is present and localize the approximate modules with embedded HT. A key challenge to this localization is that each netlist has an equal probability of being HT-infected and there can be  versatile HTs either in approximate or deterministic module(s). For simplicity of analysis, we assume the traditional side-channel analysis is effective for HT detection in the deterministic module. However, side-channel leakage can be still applied for  approximate modules. Further, we consider only combinational HT due to the limitation of the approximate synthesis tool \cite{7533498}. Different netlists would  exhibit a different amount of accuracy. Generally, the Least Significant Bit (LSB) of arithmetic modules is mostly approximated as they are error-tolerant \cite{7372600} and provide higher savings compared to approximating Most Significant Bit (MSB) of operands.   

Among $n$ netlists with no specification of error and power, a  defender wants to capitalize on the success of various techniques. As the LSB of an input word provides most approximations, an attacker would fall into embedding malicious components into LSB. As the model for approximation is unknown to the defender, the pervasiveness of error requires a defender to profile the netlist with input vector streams to determine the extent of the error. Due to heterogeneity of the netlist components, it should be possible for a defender to rank order the netlists based on error. Simultaneously, he can perform the path profiling to make it easier to determine $N$ near-critical paths from the multi-voltage multi-mode analysis. During approximation, many  near-critical paths show timing error due to voltage scaling. Hence, the paths from the slack distribution that do not violate timing constraints  are possible sources of HT. Once these paths are found, one can easily  find the modules that contain such path(s). The localization of these modules can be further examined to determine the nature of the module (approximate versus deterministic). However, it should be noted that if the HT is present in both types of modules, it will make it harder to detect the presence and timing of HT triggering signal.

\section{Experimental Results}
\label{sec:result}

To stage an attack in an approximating computing system, we first evaluate the HT vulnerability of approximate adders \cite{7167270} and approximate multipliers \cite{7827657,5718826} as mentioned in Section \ref{sec:attack}. The accuracy of individual adder and multiplier architecture in the literature \cite{7167270, 7827657,5718826} which we use in Eqn. \ref{eq:leak_delay_cost}. Next, we simulate the design under 1000 correlated input vectors \cite{8607170} to determine the power profile of an individual module.  For the same input streams, we use Synopsys VCS \cite{vcs} to calculate the rare triggering nets from the SAIF (Switching Activity Interchange Format) file. To use this information during synthesis, we use Synopsys Library Compiler \cite{lc} to build a database of HT vulnerable approximate blocks. Then both HT-free and HT infected module architectures are used as input to a modified version of ABACUS \cite{7533498}. We generate ten different netlists of a design based on Pareto fronts. We evaluate the proposed attack for two designs ({\tt FFT} and {\tt FIR}) available in ABACUS. The attack goal for both DSP cores is to transfer the filter coefficients to the outputs when a particular input sequence occurs. We have observed that with power overhead < 2\%, we are able to determine the coefficients without introducing any additional hardware errors. 

During detection, we simulate  each netlist under random and correlated input streams. After the simulation, we get the error (\%) that an approximate netlist can tolerate. We sort the netlists based on the ascending order of error (\%). For higher confidence, we perform static timing analysis (STA) using Synopsys Primetime. We extract the paths that do not violate the given timing constraints (here 10ns), and we locate the modules (approximate or deterministic) crossing these paths. We further generate test vectors to test the error resilience of the individual module. The modules having  the lowest resilience are confirmed as HT vulnerable. 

Similar to attack framework, we apply the above-mentioned detection technique and identify the rare nets responsible for leaking the coefficients in designs ({\tt FFT} and {\tt FIR}). We  detect the HT vulnerable nets and modules with an average accuracy of 90\% (false positive is 8\%, and false-negative is 2\%).

\section{Conclusion and Future Work}
\label{sec:conclude}

We present an attack and defense framework for robust modifications and detect such modifications on approximate computing synthesis. During the modifications, an attacker ranks  the available approximate modules based on accuracy, power, and rare activity nets. On the contrary, during detection, a defender utilizes the input vectors to characterize the approximate modules followed by robust path profiling. Both frameworks can be integrated seamlessly in regular design automation flow without detailed views of lower-level approximate computing circuits.  Finally, we use open source approximate modules and approximate synthesis tool to perform attack and defense. In the future, we plan to extend the extending the synthesis tool to include security processing features based on micro-architecture parameters of approximate computing.

% With only knowledge of the design, a designer can find possible HT locations easily without any expensive simulation or side-channel analysis. We also present an algorithm to rank the parameters with  lower time complexity. We describe the feasibility of bottom-up analysis approach. However, given enough run-time and good heuristics, top-down analysis is also possible. Future work includes the same parameters review for sequential HT and at a higher abstraction level (e.g., algorithmic level).
%\input{6_Definitions.tex}

\bibliographystyle{unsrt}
\scriptsize{
\bibliography{HT.bib,pba_ht.bib}

\begin{thebibliography}{10}

\bibitem{Jiang:2015:CRE:2742060.2743760}
Honglan Jiang, Jie Han, and Fabrizio Lombardi.
\newblock A comparative review and evaluation of approximate adders.
\newblock In {\em Proceedings of the 25th Edition on Great Lakes Symposium on
  VLSI}, GLSVLSI '15, pages 343--348, New York, NY, USA, 2015. ACM.

\bibitem{7167270}
M.~{Shafique}, W.~{Ahmad}, R.~{Hafiz}, and J.~{Henkel}.
\newblock A low latency generic accuracy configurable adder.
\newblock In {\em 2015 52nd ACM/EDAC/IEEE Design Automation Conference (DAC)},
  pages 1--6, June 2015.

\bibitem{8060326}
M.~A. {Hanif}, R.~{Hafiz}, O.~{Hasan}, and M.~{Shafique}.
\newblock Quad: Design and analysis of quality-area optimal low-latency
  approximate adders.
\newblock In {\em 2017 54th ACM/EDAC/IEEE Design Automation Conference (DAC)},
  pages 1--6, June 2017.

\bibitem{7372600}
S.~{Hashemi}, R.~I. {Bahar}, and S.~{Reda}.
\newblock {DRUM: A Dynamic Range Unbiased Multiplier for approximate
  applications}.
\newblock In {\em 2015 IEEE/ACM International Conference on Computer-Aided
  Design (ICCAD)}, pages 418--425, Nov 2015.

\bibitem{5718826}
P.~{Kulkarni}, P.~{Gupta}, and M.~{Ercegovac}.
\newblock Trading accuracy for power with an underdesigned multiplier
  architecture.
\newblock In {\em 2011 24th Internatioal Conference on VLSI Design}, pages
  346--351, Jan 2011.

\bibitem{opencores}
{Medium}.
\newblock
  \url{https://medium.com/zetta-venture-partners/new-strategy-in-computing-approximate-computing-1c72065ead07}.

\bibitem{Agarwal2009UsingCP}
A~Agarwal, M.~C. Rinard, S.~Sidiroglou, S.~Misailovic, and H.~Hoffmann.
\newblock {Using Code Perforation to Improve Performance, Reduce Energy
  Consumption, and Respond to Failures}.
\newblock 2009.

\bibitem{7459495}
D.~{Palomino}, M.~{Shafique}, A.~{Susin}, and J.~{Henkel}.
\newblock {Thermal optimization using adaptive approximate computing for video
  coding}.
\newblock In {\em 2016 Design, Automation Test in Europe Conference Exhibition
  (DATE)}, pages 1207--1212, March 2016.

\bibitem{6493641}
H.~{Esmaeilzadeh}, A.~{Sampson}, L.~{Ceze}, and D.~{Burger}.
\newblock {Neural Acceleration for General-Purpose Approximate Programs}.
\newblock In {\em 2012 45th Annual IEEE/ACM International Symposium on
  Microarchitecture}, pages 449--460, Dec 2012.

\bibitem{7298218}
S.~{Venkataramani}, A.~{Ranjan}, K.~{Roy}, and A.~{Raghunathan}.
\newblock {AxNN: Energy-efficient neuromorphic systems using approximate
  computing}.
\newblock In {\em 2014 IEEE/ACM International Symposium on Low Power
  Electronics and Design (ISLPED)}, pages 27--32, Aug 2014.

\bibitem{8885006}
R.~P. {Challa}, S.~{Ariful Islam}, and S.~{Katkoori}.
\newblock {An SR Flip-Flop based Physical Unclonable Functions for Hardware
  Security}.
\newblock In {\em 2019 IEEE 62nd International Midwest Symposium on Circuits
  and Systems (MWSCAS)}, pages 574--577, Aug 2019.

\bibitem{8357321}
S.~A. {Islam} and S.~{Katkoori}.
\newblock {High-level synthesis of key based obfuscated RTL datapaths}.
\newblock In {\em 2018 19th International Symposium on Quality Electronic
  Design (ISQED)}, pages 407--412, March 2018.

\bibitem{8839352}
V.~{Laguduva}, S.~A. {Islam}, S.~{Aakur}, S.~{Katkoori}, and R.~{Karam}.
\newblock Machine learning based iot edge node security attack and
  countermeasures.
\newblock In {\em 2019 IEEE Computer Society Annual Symposium on VLSI
  (ISVLSI)}, pages 670--675, July 2019.

\bibitem{Lee:2012:HES:2382196.2382323}
R.~Lee, S.~Sethumadhavan, and G.~E. Suh.
\newblock {Hardware Enhanced Security}.
\newblock In {\em Proceedings of the 2012 ACM Conference on Computer and
  Communications Security}, CCS '12, pages 1052--1052, New York, NY, USA, 2012.
  ACM.

\bibitem{8105821}
S.~{Mazahir}, O.~{Hasan}, and M.~{Shafique}.
\newblock Adaptive approximate computing in arithmetic datapaths.
\newblock {\em IEEE Design Test}, 35(4):65--74, Aug 2018.

\bibitem{7533498}
K.~{Nepal}, S.~{Hashemi}, H.~{Tann}, R.~I. {Bahar}, and S.~{Reda}.
\newblock Automated high-level generation of low-power approximate computing
  circuits.
\newblock {\em IEEE Transactions on Emerging Topics in Computing}, 7(1):18--30,
  Jan 2019.

\bibitem{6241596}
S.~{Venkataramani}, A.~{Sabne}, V.~{Kozhikkottu}, K.~{Roy}, and
  A.~{Raghunathan}.
\newblock {SALSA: Systematic logic synthesis of approximate circuits}.
\newblock In {\em DAC Design Automation Conference 2012}, pages 796--801, June
  2012.

\bibitem{Lee:2017:HSA:3130379.3130421}
S.~Lee, L.~K. John, and A.~Gerstlauer.
\newblock {High-level Synthesis of Approximate Hardware Under Joint Precision
  and Voltage Scaling}.
\newblock In {\em Proceedings of the Conference on Design, Automation \& Test
  in Europe}, DATE '17, pages 187--192, 2017.

\bibitem{Regazzoni:2018:SDS:3240765.3243497}
F.~Regazzoni, C.~Alippi, and I.~Polian.
\newblock {Security: The Dark Side of Approximate Computing?}
\newblock In {\em Proceedings of the International Conference on Computer-Aided
  Design}, ICCAD '18, pages 44:1--44:6, New York, NY, USA, 2018. ACM.

\bibitem{Yellu:2019:STA:3299874.3319453}
P.~Yellu, N.~Boskov, M.~A. Kinsy, and Q.~Yu.
\newblock {Security Threats in Approximate Computing Systems}.
\newblock In {\em Proceedings of the 2019 on Great Lakes Symposium on VLSI},
  GLSVLSI '19, pages 387--392, New York, NY, USA, 2019. ACM.

\bibitem{7342387}
N.~{Fern}, S.~{Kulkarni}, and K.~T. {Cheng}.
\newblock {Hardware Trojans hidden in RTL don't cares — Automated insertion
  and prevention methodologies}.
\newblock In {\em 2015 IEEE International Test Conference (ITC)}, pages 1--8,
  Oct 2015.

\bibitem{996017}
K.~{Deb}, A.~{Pratap}, S.~{Agarwal}, and T.~{Meyarivan}.
\newblock {A fast and elitist multiobjective genetic algorithm: NSGA-II}.
\newblock {\em IEEE Transactions on Evolutionary Computation}, 6(2):182--197,
  April 2002.

\bibitem{8342270}
J.~{Cruz}, Y.~{Huang}, P.~{Mishra}, and S.~{Bhunia}.
\newblock An automated configurable trojan insertion framework for dynamic
  trust benchmarks.
\newblock In {\em 2018 Design, Automation Test in Europe Conference Exhibition
  (DATE)}, pages 1598--1603, March 2018.

\bibitem{1585245}
L.~H. {Goldstein} and E.~L. {Thigpen}.
\newblock Scoap: Sandia controllability/observability analysis program.
\newblock In {\em 17th Design Automation Conference}, pages 190--196, June
  1980.

\bibitem{7827657}
S.~{Rehman}, W.~{El-Harouni}, M.~{Shafique}, A.~{Kumar}, J.~{Henkel}, and
  J.~{Henkel}.
\newblock Architectural-space exploration of approximate multipliers.
\newblock In {\em 2016 IEEE/ACM International Conference on Computer-Aided
  Design (ICCAD)}, pages 1--8, Nov 2016.

\bibitem{8607170}
S.~A. {Islam}, L.~K. {Sah}, and S.~{Katkoori}.
\newblock Empirical word-level analysis of arithmetic module architectures for
  hardware trojan susceptibility.
\newblock In {\em 2018 Asian Hardware Oriented Security and Trust Symposium
  (AsianHOST)}, pages 109--114, Dec 2018.

\bibitem{vcs}
{Synopsys VCS}.
\newblock \url{https://www.synopsys.com/verification/simulation/vcs.html}.

\bibitem{lc}
{Synopsys Library Compiler}.
\newblock
  \url{https://www.synopsys.com/verification/virtual-prototyping/saber/capabilities/modeling-tools.html}.

\end{thebibliography}
}
\end{document}